\documentclass[twocolumn,aps,pra,showpacs,floatfix,superscriptaddress]{revtex4}

\usepackage{epsf}
\usepackage{amsmath}
\usepackage{amsfonts}
\usepackage{amssymb}
\usepackage{bm}
\usepackage{bbm}
\usepackage{nicefrac}
\usepackage{xcolor}
\usepackage{pifont}
\usepackage{graphicx}
\usepackage{enumerate}
\usepackage{dcolumn}
\usepackage{maybemath}

\def\dd{{\mathrm{d}}}
\def\ii{{\mathrm{i}}}
\def\ee{{\mathrm{e}}}
\def\calO{{\mathcal{O}}}
\def\calZ{{\mathcal{Z}}}

\def\v{\vec}

\def\tfrac#1#2{ {\textstyle{\frac{#1}{#2}} } }

\begin{document}

\title{Non--Contact Friction for Ion--Surface Interactions}

\author{U. D. Jentschura}

\affiliation{Department of Physics, Missouri University of Science
and Technology, Rolla, Missouri 65409, USA}

\author{G. \L{}ach}

\affiliation{International Institute of Molecular and Cell Biology,
Ksi\c{e}cia~Trojdena 4, 02--109 Warsaw, Poland}

\affiliation{Faculty of Physics, University of Warsaw,
Pasteura 5, 02--093 Warsaw, Poland}

\begin{abstract}
Non-contact friction forces are exerted on physical systems through dissipative
processes, when the two systems are not in physical contact with each other,
or, in quantum mechanical terms, when the overlap of their wave functions is
negligible.  Non-contact friction is mediated by the exchange of virtual quanta,
with the additional requirement that the scattering process needs to have an
inelastic component. For finite-temperature ion-surface interactions, the
friction is essentially caused  by Ohmic resistance due to the motion of the
image charge moving in a dielectric material. A conceivable experiment is
difficult because the friction force needs to be isolated from the interaction
with the image charge, which significantly distorts the ion's flight path.  We
propose an experimental setup which is designed to minimize the influence of
the image charge interaction though a compensation mechanism, and evaluate the
energy loss due to non-contact friction for helium ions (He$^+$) interacting with
gold, vanadium, titanium and graphite surfaces. Interactions with the infinite
series of mirror charges in the plates are summed in terms 
of the logarithmic derivatives of the Gamma function, and of the 
Hurwitz zeta function.
\end{abstract}


\maketitle


%
%
\section{Introduction}
\label{sec1}

If an ion trajectory is parallel and
close to a surface (close to a ``half-space filled with a conducting
material''), then it is intuitively obvious that the motion of the image charge
inside the conductor will lead to some sort of Ohmic heating. Unless one can
show that the motion of the image charge is frictionless, this Ohmic heating
can only be compensated by a corresponding loss in the kinetic energy of the
projectile ion.  Here, we devote special attention to a rederivation of the
non-contact friction in ion-surface interactions, with an emphasis on a proposed
experimental setup which serves to eliminate the ``signal'' (the quantum
friction) from the ``noise'' (the interaction with the image charge).  The
latter can otherwise lead to a net gain or loss of kinetic energy for the ion.

Provided one accepts the existence of non-contact friction for ion-surface
interactions, the same effect (but at a smaller scale) can be expected for
atom-surface interactions. Namely, any atom constitutes, due to its quantum
mechanical nature, an ``oscillating dipole'' whose frequency-dependent response
to exciting radiation is described by the (dynamic dipole) polarizability.
Because the dipole moment fluctuates, there is a fluctuating image charge
moving in the medium, which also leads to Ohmic heating and by consequence,
non-contact friction.  However, the friction force in this case will be much
smaller, numerically, than in the case of ion-surface interactions.

One can draw a distant analogy to ion-atom versus atom-atom interactions.  It
is well known that the ion-atom interaction (or more generally, the interaction of a
charged particle with an atom) has the functional form
$1/R^4$ where $R$ is the ion-atom distance. For two neutral atoms,
the dominant term in the non-retardation limit is the van-der-Waals
interaction, which gives rise to a $1/R^6$-interaction.  For the quantum
friction effect, with an ion-surface distance $\calZ$, the damping coefficient
$\eta$ in the friction force $F = -\eta \, v$ is proportional to $1/\calZ^3$
for ion-surface interactions, which is modified to a $1/\calZ^5$-law for
atom-surface interactions~\cite{ToWi1997}. 
The pattern is clear: There are two more powers of
the distance ($R$ viz.~$\calZ$) in the denominator for interactions with
atoms, as opposed to the corresponding interaction involving an ion ``at the
other end''.

Alternatively, we may treat the case of finite
temperature~\cite{RyKrTa1989vol1,RyKrTa1989vol2,RyKrTa1989vol3,GaZo2004,Ga1991qn}
as follows~\cite{DzLiPi1959jetp,DzLiPi1961spu,ToWi1997,VoPe2002}.  When two
physical systems are in contact, then the charge fluctuations in the one will
induce mirror charge fluctuations in the other. At zero temperature, the
alteration of the available modes for quantum fluctuations due to the presence
of the surface (due to the ``boundary conditions'') gives rise to atom-wall
interactions of the Casimir--Polder type (see Chap.~8 of Ref.~\cite{Mi1994}).
At finite temperature, one has to replace the integral over the virtual photon
frequency by a summation over Matsubara frequencies [see Eq.~(30) of
Ref.~\cite{ToWi1997}].  The zero--temperature limit is obtained by the
replacement $\beta^{-1} \, \sum_{n = -\infty}^\infty \to \hbar
\int_{-\infty}^\infty \dd \omega/(2\pi)$, where the $n$th Matsubara frequency
is $\omega_n = 2 \pi/(\hbar \beta)$, and $\beta = 1/(k_B T)$ is the Boltzmann
factor. The same observation is made in Sec.~81 of Ref.~\cite{PiLi1958vol9},
where it is shown that in the zero-temperature and short-distance limit, the
Casimir  force between two solids reduces to an expression which is in
agreement with the van-der-Waals force between atoms [limit of two dilute
media, see Eqs.~(81.1), (81.9) and~(82.3) ff.~of Ref.~\cite{PiLi1958vol9}]. 
The calculation of
static forces is most easily accomplished in the imaginary-time formalism,
which leads to the Matsubara frequencies, while the calculation of friction
forces is usually done using the real-time retarded fluctuation-dissipation
theorem~\cite{ToWi1997,VoPe2002}.  Indeed, in Ref.~\cite{VoPe2002}, it is shown
that the friction force on an atom in the vicinity of a dielectric surface can
be obtained from a calculation of the van-der-Waals interaction for an atom
which undergoes a small periodic mechanical oscillation, after subtracting the
conservative static, and the conservative oscillatory 
component (vibration component) of
the van-der Waals force [see Eq.~(24) of Ref.~\cite{VoPe2002}].  

In order to calculate the friction force, one has to describe the correlation
of electric field fluctuations at different points, in the presence of a
dielectric material filling the half-space $z < 0$, due to charge fluctuations
in the
dielectric~\cite{RyKrTa1989vol1,RyKrTa1989vol2,RyKrTa1989vol3,PiLi1958vol9}.
We here use the real-time formulation of the fluctuation-dissipation theorem
and apply it from first principles.  Another point is that one has to apply the
Wick theorem to the thermal fluctuations, remembering that the Wick theorem
holds both for quantum as well as for thermal (statistical) fluctuations.
Finally, the ``thermal factors'' need to be taken into account accurately, with
a full account of the quantum nature of the problem.  Last, but certainly not
least, one has to be careful in applying the conventions for the Fourier
transform in the theory of thermal fluctuations and in electromagnetic signal
theory~\cite{RyKrTa1989vol1,RyKrTa1989vol2,RyKrTa1989vol3} correctly.  These
can otherwise lead to inconsistent prefactors in the final results.

We continue in Sec.~\ref{sec2} with a discussion of basic concepts underlying
the fluctuation-dissipation theorem, which is central to the derivation of the
finite-temperature non-contact friction effect, and with the thermal
correlations of the electric field.  The ion-surface interaction and the
calculation of the non-contact friction force on a charged particle are
discussed in Sec.~\ref{sec3}.  Finally, a sketch of a proposed experimental
setup is discussed in Sec.~\ref{sec4}. It involves two parallel conducting
plates which give rise to a series of mirror charges and requires the
calculation of interaction potentials with the mirror charges.
Conclusions are drawn in Sec.~\ref{sec5}.

%
%
\section{Fluctuation--Dissipation Theorem} \label{sec2}

%
%
\subsection{Fluctuation--Dissipation Theorem}

We start by recalling the Boltzmann factor $\beta$ and the 
bosonic thermal occupation number $n(\omega)$,
\begin{equation}
\beta = \frac{1}{k_B \, T} \,, 
\qquad
n(\omega) = \frac{1}{\exp(\beta\, \hbar\, \omega) - 1} \,.
\end{equation}
The Kallen--Welton thermal factor $\Theta(\omega, T)$ is given by the 
relation
\begin{equation}
\label{KallenWelton}
\Theta(\omega, T) = \hbar \, \omega \,
\left( \tfrac12 + n(\omega) \right)
= \tfrac12 \hbar\, \omega \,
\coth\left( \tfrac12 \, \hbar\, \beta \, \omega \right) \,.
\end{equation}
It has the properties,
\begin{subequations}
\begin{align}
\label{Thetasym}
\Theta(\omega, T) =& \; -\Theta(-\omega, T)  \,,
\\[0.133ex]
\label{relation1}
\left( 1 + n(\omega) \right) \,
\left( 1 + n(-\omega) \right) = & \;
\frac{1}{\hbar\, \beta} \, 
\frac{\partial n(\omega)}{\partial \omega} \,,
\\[0.133ex]
\left( 1 + n(\omega) \right) \, n(\omega) =& \;
-\frac{1}{\hbar\, \beta} \,
\frac{\partial n(\omega)}{\partial \omega} \,.
\end{align}
\end{subequations}
In the high-temperature limit, we have
$\Theta(\omega, T) \to 1/\beta = k_B \, T$.
Let now $x$ be an observable of a dynamical system with Hamiltonian 
$H_0(x)$ subject to thermal fluctuations.
We assume that $x(t)$ fluctuates around its mean value 
$\langle x \rangle_0$ with fluctuations characterized by 
a power spectrum 
\begin{equation}
S_x(\omega) = \left< x(t) \, x(0) \right>_\omega 
= \int \frac{\dd \omega}{2\pi} \, \ee^{-\ii \, \omega \, t} \, 
\left< x(t) \, x(0) \right> \,.
\end{equation}
Let us consider the conjugate variable of $x$,
namely, a scalar force field $f$ which alters the Hamiltonian to 
\begin{equation}
\label{defHfull}
H = H_0(x) + x(t) \, f(t) \,.
\end{equation}
The response of an observable $x$ to the field 
term is characterized (to first order) 
by the susceptibility of the linear response function 
$\chi(t)$ of the system, 
\begin{equation}
\label{suscep}
\langle x(t) \rangle = \langle x \rangle_0 +
\int\limits_{-\infty}^t \chi(t - \tau) \, f(\tau) \, \dd \tau \,,
\end{equation}
where the mean value $\langle x \rangle_0$
is obtained as a thermal average taken over the distribution
governed by the unperturbed Hamiltonian $H_0(x)$,
i.e., with respect to the weight function
\begin{equation}
W_0(x) = \frac{\exp\left[ - \beta \, H_0(x) \right]}%
{\int \dd x' \, \exp\left[ - \beta \, H_0(x') \right]} \,.
\end{equation}
The field term is adiabatically switched on at 
$\tau = -\infty$.
Expanding the weight function of the full Hamiltonian
given in Eq.~\eqref{defHfull} for small $f(t)$, 
one can easily motivate~\cite{wikifluc} that the
the imaginary part of the complex susceptibility
\begin{equation}
\chi(\omega) = 
{\rm Re}\left[ \chi(\omega) \right] +
\ii \, {\rm Im}\left[ \chi(\omega) \right] \,,
\end{equation}
is related to the power spectrum of fluctuations, which
by the fluctuation-dissipation 
theorem~\cite{RyKrTa1989vol1,RyKrTa1989vol2,RyKrTa1989vol3,GaZo2004,Ga1991qn}
reads as
\begin{equation}
\label{Sx}
\langle x(t) \, x(0) \rangle_\omega = S_x(\omega) 
= \frac{2 \Theta(\omega, T)}{\omega} \;
{\rm Im}\left[\chi(\omega) \right] \,.
\end{equation}
This involves the Kallen--Welton thermal factor.
In the high-temperature limit, one may 
replace $\Theta(\omega, T) \to 1/\beta = k_B \,T$.
We should note that 
the conventions for the Fourier transform
used in the current work follow those commonly used in physics,
\begin{equation}
{\widetilde f}(\omega) = \int \dd t\, \ee^{\ii \, \omega \, t} \, f(t) \,,
\quad
f(t) = \int \frac{\dd \omega}{2\pi} \, \ee^{-\ii \, \omega \, t} \, 
{\widetilde f}(\omega) \,,
\end{equation}
and these are different from those in electrical engineering
(see Refs.~\cite{RyKrTa1989vol1,RyKrTa1989vol2,RyKrTa1989vol3}).

%
%
\subsection{Correlation Function for the Electric Field}

Let us try to motivate a formula
for the thermal correlation of the electric 
field in the vicinity of a dielectric material,
based on the thermal charge fluctuations inside the 
dielectric. We first observe that the 
scalar potential $\Phi(\vec r)$ is given as 
\begin{equation}
\Phi(\vec r) = \int \dd^3 r' \, G(\vec r, \vec r') \,
\rho(\vec r') \,,
\end{equation}
where $\rho$ is the charge density.
In free space, the Green function is given as 
$G(\vec r, \vec r ') = 
\left( 4 \pi \epsilon_0 | \vec r - \vec r'| \right)^{-1}$.
In the vicinity of a dielectric wall, one has 
to modify the Green function as follows (non-retardation limit,
see Ref.~\cite{ToWi1997}),
\begin{align}
\label{GGG}
G(\vec r, \vec r ') \to & \; G(\omega, \vec r, \vec r ') = 
\frac{1}{4 \pi \epsilon_0 |\v{r}'-\v{r}|} 
\nonumber\\[0.133ex]
& - \frac{\epsilon(\omega)-1}{\epsilon(\omega)+1}\,
\frac{1}{4 \pi \epsilon_0 |\v{r}'-\v{r} + 
2 \v{n}_\perp(\v{r}\cdot\v{n}_\perp)|} \,.
\end{align}
We here ignore a possible path difference of the emitted perturbation (at $\vec
r$) and the incoming wave (at $\vec r'$). The normal vector 
$\vec n_\perp$ is the outward normal pointing away from the 
half-space filled with the dielectric material. Indeed, we shall need the Green
function in the limit $\vec r \to \vec r'$. In this limit, the path difference
of the reflected and the emitted wave vanishes.  Corrections to this result due
to the velocity of the atom in the $x$ direction are of order $\calZ \omega
v_x^2/c^3$ and are negligible on the order of interest for the current paper.
Here, $\calZ$ denotes the ion-surface distance.

Our formula~\eqref{GGG} is in agreement with Eq.~(6) of Ref.~\cite{ToWi1997},
which gives the fluctuations of the scalar instead of the vector potential, due
to charge fluctuations in the vicinity of a dielectric wall.  The $1/(2\pi)$
prefactor in Eq.~(6) of Ref.~\cite{ToWi1997} is due to the different Fourier
transform conventions used therein, which follow the conventions commonly
adopted in electrical engineering. The
correlation function of the fluctuations of the scalar potential reads, in full
agreement with the general paradigm set by the fluctuation-dissipation 
theorem given in Eq.~\eqref{Sx},
\begin{equation}
\left< \Phi(\vec r) \, \Phi(\vec r') \right>_\omega =
\frac{2 \Theta(\omega, T)}{\omega} \,
{\rm Im} \left[ G(\omega; \vec r, \vec r') \right] \,.
\end{equation}
Here, $\rho$ is the ``fluctuating force'',
whereas $\Phi(\vec r)$ takes the role of the fluctuating signal.
The correlation function for the electric field follows 
by differentiation,
\begin{align}
\left< E_i(\vec r) E_j(\vec r') \right>_\omega =& \;
-\left< \nabla_i \Phi(\vec r) \nabla'_j \Phi(\vec r') \right>_\omega 
\\[0.133ex]
=& \; -\frac{2 \Theta(\omega, T)}{\omega} \,
{\rm Im} \left[ \nabla_i \nabla'_j G(\omega; \vec r, \vec r') \right] \,,
\nonumber\\[0.133ex]
=& \; \frac{2 \, \Theta(\omega, T)}{\omega} \,
{\rm Im} \left( \frac{\epsilon(\omega)-1}{\epsilon(\omega)+1} \right)  
\nonumber\\[0.133ex]
& \; \times \nabla_i \nabla'_j 
\frac{1}{4 \pi \epsilon_0 |\v{r}'-\v{r}+ 2 \v{n}_\perp(\v{r}\cdot\v{n}_\perp)|} \,.
\nonumber
\end{align}
The minus sign is explained because negatively correlated
charge fluctuations at two points along a reference line 
generate fluctuations of the electric field
which are positively correlated along the line joining the 
two points. Here, $\epsilon(\omega)$
denotes the relative dielectric function
(relative permittivity),
which has to be multiplied by
the vacuum permittivity $\epsilon_0$,
if one would like to obtain the full electric 
permittivity of the medium.

There is one last subtlety to discuss.
We should be careful interpreting the correlation function
$\left< E_i(\vec r) E_j(\vec r') \right>_\omega$.
Namely, according to Eqs.~(32) and~(49) of Ref.~\cite{ToWi1997}, it is 
more accurate to relate the following, symmetrized 
correlation function to the imaginary part of the 
susceptibility, according to the replacement
%
\begin{align}
& \left< E_i(\vec r) E_j(\vec r') \right>_\omega \to
\int_{-\infty}^\infty \dd t \,
\cos(\omega \, t) \,
\, \left< E_i(\vec r, t) E_j(\vec r', 0) \right>
\nonumber\\[0.133ex]
&  = \tfrac12 \,
\int_{-\infty}^\infty \dd t \,
\left( \ee^{\ii \omega t} + \ee^{-\ii \omega t} \right) \,
\, \left< E_i(\vec r, t) E_j(\vec r', 0) \right>
\nonumber\\[0.133ex]
&
= \tfrac12 \, \int_{-\infty}^\infty \dd t \, \ee^{\ii \omega t}
\, \left< E_i(\vec r, t) E_j(\vec r', 0) \right>
\nonumber\\[0.133ex]
& \qquad + \tfrac12 \, \int_{-\infty}^\infty \dd t \, \ee^{\ii \omega t}
\, \left< E_i(\vec r, -t) E_j(\vec r', 0) \right>
\nonumber\\[0.133ex]
&
= \tfrac12 \, \int_{-\infty}^\infty \dd t \, \ee^{\ii \omega t}
\, \left< \{ E_i(\vec r, t), E_j(\vec r', 0) \} \right>
\nonumber\\[0.133ex]
&
= \left< \tfrac12 \{ E_i(\vec r, t), E_j(\vec r', 0) \} \right>_\omega \,.
\end{align}
%
The distinction between 
$\left< E_i(\vec r) E_j(\vec r') \right>_\omega$ 
and
$\left< \tfrac12 \{ E_i(\vec r, t), E_j(\vec r', 0) \} \right>_\omega$
becomes important when more than two operators are 
involved~\cite{ToWi1997,JeLaJaDK2015atom}.
So, we should write, more correctly,
\begin{align}
\label{EEG}
& \left< \tfrac12 \{ E_i(\vec r, t), E_j(\vec r', 0) \} \right>_\omega 
= \frac{2 \, \Theta(\omega, T)}{\omega} \,
{\rm Im} \!\left[ \frac{\epsilon(\omega)-1}{\epsilon(\omega)+1} \right] \!
\nonumber\\[0.133ex]
& \qquad \times 
\nabla_i \nabla'_j
\frac{1}{4 \pi \epsilon_0 |\v{r}'-\v{r}+ 2 \v{n}_\perp(\v{r}\cdot\v{n}_\perp)|}.
\end{align}
The result in Eq.~\eqref{EEG} is consistent with 
Eq.~(15) of Ref.~\cite{AnPiStSv2008}.

%
%
\section{Quantum Friction for Ions}
\label{sec3}

%
%
\subsection{Ion--Surface Interactions}

Armed with the results from Sec.~\ref{sec2}, we are now in the position to
evaluate the friction force due to the mirror charge running in the dielectric,
which generates Ohmic heating.  The force on a charged particle is given by 
\begin{equation}
\vec F(t) = Z \, e \, \vec E(\vec r, t) \,,
\end{equation}
where we assume $e$ to denote the electron charge, 
and the ion is $Z$-fold negatively charged.
We also use the result 
[$\vec r = (x,y,\calZ)$ and $\vec r' = (x',y',\calZ')$]
\begin{align}
\label{alg1}
& \lim_{\v{r}'\to\v{r}}
\nabla_x
\nabla'_x
\frac{1}{|\v{r}'-\v{r}+2 \, \v{n}_\perp(\v{r}\cdot\v{n}_\perp)|}
= \frac{1}{8 \, \calZ^3} \,.
\end{align}
We consider the formula for the friction force,
\begin{equation}
F_x = - \eta \, v_x\,.
\end{equation}
According to Eq.~(8.15) of Ref.~\cite{Ku1966}
(for a more modern perspective
Ref.~\cite{MaPuRoVu2008}),
the fluctuation-dissipation theorem 
determines the friction force
via a Green--Kubo formula.
In the derivation, we use the fact that 
for $ t> 0$, we have $\left< F_x(t) \, F_x(0) \right> = 
\left< F_x(0) \, F_x(-t) \right>$ 
due to time-translation invariance.
We can thus symmetrize the integrand as follows,
\begin{align}
\eta =& \; \beta \int_0^\infty \dd t\, \left< F_x(t) \, F_x(0) \right>
\nonumber\\[0.133ex]
=& \; \frac{\beta}{2} \int_{-\infty}^\infty \dd t\, 
\left< \tfrac12 \, \{ F_x(t), \, F_x(0) \} \right>
\nonumber\\[0.133ex]
=& \; \frac{\beta}{2} (Z e)^2 \, 
\lim_{\vec r \to \vec r'} \int_{-\infty}^\infty \dd t\, 
\left< \tfrac12 \, \{ E_x(\vec r, t), \, E_x(\vec r', 0) \} \right>
\nonumber\\[0.133ex]
=& \; \frac{\beta}{2} (Z e)^2 \, 
\lim_{\omega \to 0} \left\{ \frac{2 \Theta(\omega, T)}{\omega} \,
{\rm Im} \left( \frac{\epsilon(\omega)-1}{\epsilon(\omega)+1} \right)
\right\}
\nonumber\\[0.133ex]
& \; \times 
\lim_{\vec r \to \vec r'}
\nabla_x \nabla'_x
\frac{1}{4 \pi \epsilon_0 |\v{r}'-\v{r}+ 2 \v{n}_\perp(\v{r}\cdot\v{n}_\perp)|}
\nonumber\\[0.133ex]
=& \; \frac{\beta (Z e)^2}{4 \pi \epsilon_0} \,
\lim_{\omega \to 0} \left\{ \frac{1}{\beta \, \omega} 
{\rm Im} \left( \frac{\epsilon(\omega)-1}{\epsilon(\omega)+1} \right)
\right\}
\, \frac{1}{8 \calZ^3} 
\nonumber\\[0.133ex]
=& \; \frac{(Z e)^2}{32 \pi \epsilon_0} \, 
\underbrace{ \lim_{\omega \to 0} \left\{ \frac{1}{\omega} \,
{\rm Im} \left( \frac{\epsilon(\omega)-1}{\epsilon(\omega)+1} 
\right) \right\}}_{\equiv L} \, \frac{1}{\calZ^3} \,.
\end{align}
The end result is 
\begin{subequations}
\label{ion_general}
\begin{align}
\eta =& \; \frac{(Z e)^2 \, L}{32 \pi \epsilon_0 \, \calZ^3} \,,
\\[2ex]
\label{defL}
L =& \; \lim_{\omega \to 0} \left\{ \frac{1}{\omega} \,
{\rm Im} \left( \frac{\epsilon(\omega)-1}{\epsilon(\omega)+1} \right) \right\} \,.
\end{align}
\end{subequations}
We note that the drag force $F_x = -\eta \, v_x$ is 
independent of the mass of the projectile particle;
it only depends on its charge state.
The result~\eqref{ion_general} depends
on the asymptotic shape of the Boltzmann factor
$n(\omega) = 1/(\exp(\beta\hbar\omega)-1)$,
which goes as $1/(\beta\hbar\omega)$, for small $\omega$.
The final evaluation of the limit $L$ 
proceeds via a consideration of the 
low-frequency limit of the dielectric function 
of the material and crucially depends on the 
lowest electronic resonance frequency of the material.

A final word on retardation corrections is in order.
In the treatment of Casimir-Polder interactions of 
neutral atoms, the parameter governing the retardation 
effects is~\cite{Mi1994,LaDKJe2010pra}
\begin{equation}
\label{defxi}
\xi = \frac{\omega \, \calZ}{c} \,,
\qquad
\omega \sim \frac{\alpha^2 \, m_e \, c^2}{\hbar} \,,
\end{equation}
where we indicate a typical value of an atomic 
transition frequency $\omega$ 
(the fine-structure constant is $\alpha$).
Retardation sets in when the phase of the atomic
oscillation changes significantly over the time 
it takes light to travel to the surface and back, i.e.,
when $\xi \sim 1$ and thus
\begin{equation}
\calZ \sim \frac{a_0}{\alpha} \approx 137 \, a_0 \,,
\qquad
a_0 = \frac{\hbar}{\alpha \, m_e \, c} \,.
\end{equation}
Retardation changes the leading $1/\calZ^3$ 
interaction for short distances to a $1/\calZ^4$ 
term at long range~\cite{Mi1994,LaDKJe2010pra}.

For the ion-surface interaction, the 
leading conservative term, for any distance, is given by 
the $1/\calZ$ attractive interaction with the mirror
charge, and, for the dissipative friction term, 
by the $1/\calZ^3$ form given in Eq.~\eqref{ion_general}.
The estimate of the retardation corrections then has to proceed 
differently; it is necessary 
to evaluate the retardation correction to the 
(in principle electrostatic) interaction of the ion with 
its mirror charge. One should compare (i) the time it takes
light to travel to the surface and back to the 
ion moving alongside and parallel to the surface, 
to (ii) the time it would light to travel back to a static ion.
The relative difference is a measure of the retardation 
correction to the electrostatic interaction.
The expansion parameter in this case is easily found to be
\begin{equation}
\xi' = \frac{v}{c} \,,
\end{equation}
where $v$ is the ion's velocity. Because these fly-by 
velocities are much smaller than $c$ in 
typical atomic-beam or ion-beam experiments, we can safely 
ignore the retardation corrections.
Further retardation effects due to the frequency of the 
exchange photon can be ignored; the 
result given in Eq.~\eqref{defL} is formulated in terms of 
a limit for small frequencies $\omega \to 0$.

\begin{figure*}
\begin{center}
\begin{minipage}{0.91\linewidth}
\includegraphics[width=0.9\linewidth]{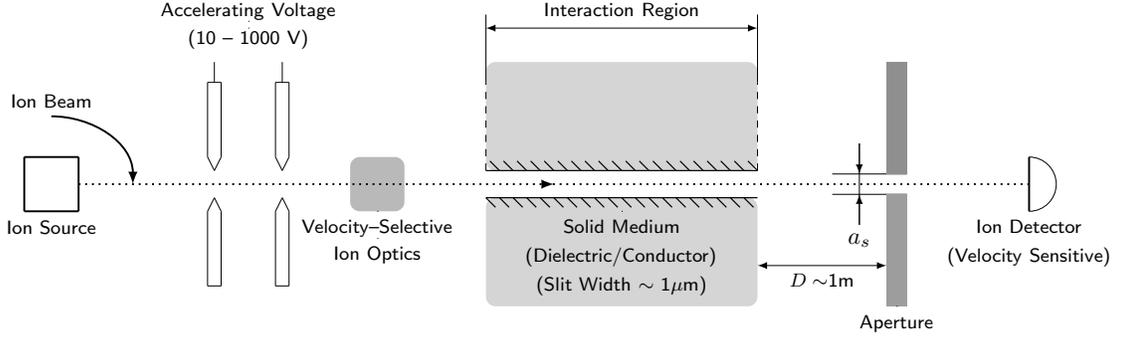}
\caption{\label{fig1} Basic proposed setup for the 
measurement of ionic non-contact friction. 
An incoming ion beam is velocity-selected and 
enters an interaction region with a compensating slit configuration,
involving two parallel bars of (preferentially single-crystal) 
material, whose induced non-contact friction is to be studied in the 
experiment. A further free beam path of a few meters in length 
serves to eliminate conceivable contributions to the 
ionic energy loss due to imperfect alignment of the 
interaction region. Finally, the energy of the 
emerging ion can be measured after it has passed
through an aperture.}
\end{minipage}
\end{center}
\end{figure*}

%
%
\subsection{Material and Friction Force}

A rather general functional form for the dielectric 
function of a material is given as~\cite{LaDKJe2010pra,DeJo2012}
\begin{equation}
\label{model_epsilon}
\epsilon(\omega) = 
1 + \sum_{n=0}^\infty \frac{a_n - \ii \, b_n \, \omega}{\omega_n^2 - \omega^2 
- \ii \, \gamma_n \, \omega} \,.
\end{equation}
The expansion coefficient $a_n$ is an amplitude which can be written as
\begin{equation}
a_n = \Delta \epsilon_n \, \omega_n^2 \,,
\qquad
b_n = \Delta \epsilon_n \, \gamma'_n \,.
\end{equation}
Recently, an approach with manifestly complex
expansion coefficients has led to an excellent fit of the 
dielectric function of silicon over a
wide frequency range~\cite{DeJo2012}.
The static limit of the dielectric function then reads as
\begin{equation}
\epsilon(\omega=0) = 
1 + \sum_{n=0}^\infty \Delta \epsilon_n \,.
\end{equation}
The functional behaviour of $\epsilon(\omega)$ 
for small $\omega$ crucially depends on the value of the 
lowest resonance frequency $\omega_{n=0}$,
where we reserve the subscript $n=0$ for the 
case of an excitation at zero resonance frequency 
$\omega_0 \equiv 0$ and otherwise set its multiplying coefficient 
to zero in the case of the absence of such a resonance.
We ascertain that if a resonance with 
$\omega_0 = 0$ exists, then $\epsilon(\omega) - 1 = \calO(\omega^{-1})$ 
for small angular frequencies.
This is the case for metals and other conductors, where,
by definition, the energy required to excited an 
electron into the conduction band vanishes.
By contrast, if $a_0 = 0$ and the lowest resonance 
frequency of the material is $\omega_1 \neq 0$, then
$\epsilon(\omega) = \epsilon(0) + \calO(\omega)$,
so that ${\rm Im}[\epsilon(\omega)] = \calO(\omega)$
for small $\omega$. In both cases, the limit $L$ 
defined in Eq.~\eqref{defL} is finite.

Let us first consider the case of a conductor.
Then, the dominant contribution for small $\omega$ is 
given by the term with $n=0$ in Eq.~\eqref{model_epsilon},
\begin{equation}
\epsilon(\omega) 
\approx 1 - \frac{a_0}{\omega (\omega + \ii \, \gamma_0)}
= 1 - \frac{\omega_p^2}{\omega (\omega + \ii \, \gamma_0)} \,,
\end{equation}
where $\omega_p$ is the plasma frequency of the 
free electron gas constituting the conduction band electrons.
This leads to 
\begin{equation}
\label{Lconduc}
L = \lim_{\omega \to 0} \left\{ \frac{1}{\omega} \,
{\rm Im} \left( \frac{\epsilon(\omega)-1}{\epsilon(\omega)+1} \right) \right\}
= \frac{2 \gamma_0}{\omega_p^2} 
= \frac{2 \epsilon_0}{\sigma_T(0)} \,,
\end{equation}
where 
\begin{equation}
\sigma_T(0) = \epsilon_0 \, \frac{\omega_p^2}{\gamma_0}
\end{equation}
is the temperature-dependent conductivity (measured in 
inverse $\Omega \, {\rm m}$) of the material, for 
a direct current (alternating current of zero frequency).
We recover the result given in Eq.~(35) of Ref.~\cite{ToWi1997}
for a conductor,
\begin{equation}
\label{ion_conductor}
\eta = \frac{(Z e)^2}{16 \, \pi \, \sigma_T(0) \, \calZ^3} \,.
\end{equation}
The only temperature-dependence remaining in this result is 
via the implicit dependence of the ``static''
conductivity $\sigma_T(0)$ on the 
temperature. In the low-temperature limit,
many materials become superconducting,
which implies that their conductivity 
$\sigma_T(0)$ diverges and the 
friction coefficient $\eta$ vanishes.

For an undoped semiconductor and other materials such as gases, 
the evaluation of the limit $L$ proceeds via 
a consideration of the lowest resonance. In the case $b_1 = 0$, 
we have
\begin{subequations}
\begin{align}
\epsilon(\omega) \approx & \;
1 + \frac{a_1}{\omega_1^2 - \omega^2 - \ii \, \gamma_1 \, \omega} \,,
\\[0.133ex]
L =& \; \frac{2 \, a_1 \, \gamma_1}{(a_1 + 2 \omega_1^2)^2} \,,
\\[0.133ex]
\eta =& \frac{(Z e)^2 \, a_1 \, \gamma_1}%
{16 \, \pi \, (a_1 + 2 \omega_1^2)^2 \, \calZ^3} \,.
\end{align}
\end{subequations}
Here, we assume that the lowest resonance dominates,
i.e., that $\omega_2^2 \gg \omega_1^2 \gg a_1$.
If a modified Lorentz profile with $\gamma'_1$ is chosen
(according to Ref.~\cite{DeJo2012}),
then the result simplifies to 
\begin{subequations}
\label{epsmod}
\begin{align}
\epsilon(\omega) \approx & \;
1 + \sum_{n=1}^\infty 
\frac{\Delta \epsilon_n (\omega_n^2 - \ii \, \gamma'_n \, \omega)}%
{\omega_n^2 - \omega^2 - \ii \, \gamma_n \, \omega} \,,
\\[0.133ex]
L =& \; 
\sum_{n=1}^\infty \frac{2 \, \Delta \epsilon_n \, (\gamma_n - \gamma'_n)}%
{(2 + \sum_m \Delta\epsilon_m)^2 \, \omega_n^2} \,,
\\[0.133ex]
\eta =& \frac{(Z e)^2}{16 \, \pi \, \calZ^3} \,
\sum_n \frac{\Delta \epsilon_n \, (\gamma_n - \gamma'_n)}%
{(2 + \sum_m \Delta\epsilon_m)^2 \, \omega_n^2} \,,
\end{align}
\end{subequations}
where in the last line both sums extend over
$m,n = 1, \dots, \infty$.
Let us match these formulas with the known expressions for 
a dilute gas (bulk material), whose dielectric function is given by
\begin{align}
\label{epsmodalt}
\epsilon(\omega) =& \;
1 + \sum_n \frac{N_V}{\epsilon_0} \, \alpha(\omega) 
\nonumber\\[0.133ex]
=& \; 1 + \frac{N_V}{\epsilon_0} \sum_n \frac{f_{n0}}%
{E_{n0}^2 - \ii \, \Gamma_n \, (\hbar\omega) - (\hbar\omega)^2} \,.
\end{align}
Here, $N_V = N/V$ is the volume density of atoms, and the $E_{n0}$
are the excitation energies from the 
ground state to the $n$th excited state. The 
(dipole) polarizability is $\alpha(\omega)$, 
and the dipole oscillator strength for a 
transition from the reference state $k$ to the 
virtual state $n$ is $f_{nk} \equiv f^{(\ell=1)}_{nk}$,
where 
\begin{align} 
\label{oscstrengthdef}
f_{n0} = & \;
2 \, \frac{4 \pi e^2}{(2 \ell + 1)^2} \, (\hbar \, \omega_{n0}) \,
\nonumber\\[0.133ex]
& \; \times \sum_{m_k} \sum_{m}
\left| \left< \phi_k \left| 
\sum_{i} \, (r_i)^\ell \, Y_{\ell m}(\hat{r}_i) \right| \phi_0 \right> \right|^2  \,.
\end{align}
Here, $\ell=1$ denotes the dipole contribution to the 
polarizability, where $2^\ell$ is the multipole order~\cite{YaBaDaDr1996}.
We sum over the magnetic projections $m$
of the spherical harmonic and over the
magnetic projections $m_k$ of the excited state $\phi_k$.
The identification of Eqs.~\eqref{epsmod} and~\eqref{epsmodalt}
then proceeds as follows,
\begin{equation}
\omega_n = \frac{E_{n0}}{\hbar} \,,
\quad
\gamma_n = \frac{\Gamma_n}{\hbar} \,,
\quad
\gamma'_n = 0 \,,
\quad
\Delta\epsilon_n = \frac{N_V f_{n0}}{\epsilon_0 \, E^2_{n0}} \,.
\end{equation}
These quantities can directly be used in Eq.~\eqref{epsmod}.

%
\section{``Sniper'' Setup Configuration}
\label{sec4}

%
%
\subsection{Mirror Charges for Parallel Conducting Plates}
\label{mirror}

We consider an experimental setup as given in Fig.~\ref{fig1} and strive to
calculate the interaction potential of the ion with the mirror charges in both
the upper, as well as the lower slab, as well as the induced friction force.
For simplicity, we shall consider, in the following, 
two conducting parallel plates.
The upper surface of the lower plate is in the $xy$ plane ($z = 0$), whereas the
upper plate is at $z = a$.  The ion's $z$ coordinate is denoted as $\calZ$.  The
following analysis is in part inspired by Refs.~\cite{Za1976,LuRa1985}.  The
(positively charged) ion generates two image charges in the two conductors,
which in turn, by mirroring them against the respective other conducting plate,
generate an infinite series of mirror charges, which can be identified as
follows.  Iterative calculation of the positions of the mirror charges leads
to are two ``upper'' series of mirror charges, negative ones at position $2 n a
- \calZ$ [the distance to the reference point $\calZ$ is $2 (n a - \calZ)$], and
positive ones at positions $2 n a + \calZ$ (the distance is $2 n a$).  There
are also two series of mirror charges in the lower slab, positive ones at
positions $- 2 n a + \calZ$  (with a distance of $2 n a$), and negative ones at
position $-2 (n-1)a - \calZ$, whose distance to the reference point is $2 ((n-1)
a + \calZ)$.

Let us consider the limiting case of the ion being close to the lower plate at
$z=0$, i.e., the limit $\calZ \to 0$.  The dominant interaction potential will be
due to the closest mirror charge in the lower slab, which is located at
$z=-\calZ$. The distance of charge and mirror charge is $2 \calZ$, and the
interaction potential is $- (Z e)^2/(16 \pi \epsilon_0 \calZ)$, where the
prefactor takes care of the distance $2 \calZ$, as well as the fact that
the electric field is zero inside the conductor (see the Complement to Chap.~11
of Ref.~\cite{CTDiLa1978vol2}). Alternatively, the additional factor $1/2$ can
be understood as follows: If one is to move the charge upward by a given
distance, then the mirror charge moves ``automatically'', and thus the 
work required to move the charge (and its interaction potential) is halved.
The sum over all other mirror charges leads to the 
following total interaction potential,
\begin{equation}
V = - \frac{(Z e)^2}{4 \pi \epsilon_0} \, C(\calZ) \,.
\end{equation}
A series representation of the correction factor is 
as follows (again in the non-retardation limit, 
see Ref.~\cite{LuRa1985}),
\begin{align}
C(\calZ) =& \; \frac12 \, \sum_{n=1}^\infty
\left( \frac{1}{2 a (n-1) + 2 \calZ} 
+ \frac{1}{2 a n - 2 \calZ} 
- \frac{1}{a n} \right) 
\nonumber\\[0.133ex]
=& \; -\frac{1}{4 a} \, \left[ \Psi\left(\frac{\calZ}{a}\right) 
+ \Psi\left(1 - \frac{\calZ}{a} \right) 
+ 2 \, \gamma_E \right] \,,
\end{align}
where $\gamma_E = 0.57721\dots$ is the Euler--Mascheroni constant.
The logarithmic derivative of the $\Gamma$ function 
and its generalization $\Psi^{(n)}(x)$ read as follows,
\begin{equation}
\Psi(x) = \frac{\dd}{\dd x} \ln[\Gamma(x)] \,,
\qquad
\Psi^{(n)}(x) = \frac{\dd^n}{\dd x^n} \psi(x) \,.
\end{equation}
We anticipate that we shall need $\Psi^{(n=2)}(x)$ in the following
derivations. For small $\calZ$, the asymptotic expansion reads as 
\begin{equation}
C(\calZ) = \frac{1}{4 \, \calZ} + 
\frac{\zeta(3)}{2 a^3} \, \calZ^2 + \calO(\calZ^4) \,,
\end{equation}
in agreement with the consideration sketched above.

For the ``sniper'' configuration to be discussed below,
the setup sketched in Fig.~\ref{fig1} suggests to assume an
(almost) symmetric configuration, with the atom in the middle 
between the two conducting slabs, i.e., one expands about the 
point $\calZ \approx a/2$. 
For perfect symmetry, we have
\begin{equation}
C\left(\calZ=\frac{a}{2}\right) = 
\frac{\ln(2)}{a} =
\frac{0.693147}{a} < \frac{1}{a} \,.
\end{equation}
If we assume perfect symmetry, then the 
two nearest mirror charges are at a distance 
$a$ from the ion, which, together with the 
correction factor $1/2$ of Ref.~\cite{CTDiLa1978vol2},
would suggest a value of $2 \times 1/(2 a) = 1/a$
for the $C$ coefficient. The contribution of the remaining mirror 
charges reduces this result to the value $\ln(2)/a$.

We now consider the contribution of the 
remaining mirror charges to the friction.
To this end, we first recall the result from Eq.~\eqref{ion_general},
which is initially valid for a single surface,
\begin{equation}
\label{matching}
\eta = \frac{(Z e)^2 \, L}{32 \pi \epsilon_0 \, \calZ^3} \; \to \;
\eta_{(2)} = \frac{(Z e)^2 \, L}{32 \pi \epsilon_0} \, D(\calZ) \,.
\end{equation}
Here, the variable $\calZ$ in the 
expression for $\eta$ denotes the distance 
of the ion and the conducting wall in the 
case of the presence of a single wall.
This result will need to be matched against
a function $D(\calZ)$, which measures the effect
in the case of two parallel slabs
and sums over all mirror charges,
with the proviso that the interaction here is
proportional to the third inverse power of the distance.
A series representation of the correction factor is 
easily obtained as follows,
\begin{align}
\label{resD}
D(\calZ) =& \; \sum_{n=1}^\infty
\left( \frac{1}{[a (n-1) + \calZ]^3} 
+ \frac{1}{(a n - \calZ)^3} 
- \frac{2}{(a n)^3} \right) 
\nonumber\\[0.133ex]
=& \; -\frac{1}{2 \, a^3} \, 
\left[ \Psi^{(2)}\left(\frac{\calZ}{a}\right) + 
\Psi^{(2)}\left(1-\frac{\calZ}{a}\right) + 4 \, \zeta(3) \right] 
\nonumber\\[0.133ex]
=& \; \frac{1}{\calZ^3} + 
\frac{12 \, \zeta(5)}{a^3} + \calO(\calZ^2) \,.
\end{align}
The latter expansion confirms the consistency with 
the result for a single wall, as given in Eq.~\eqref{matching}.
The result given in Eq.~\eqref{resD}
can be expressed in terms of the  Hurwitz generalized zeta function,
\begin{equation}
\zeta(b,x) = \sum_{n=0}^\infty \frac{1}{(n+x)^b} \,,
\qquad
\zeta(3,x) = - \frac12 \, \psi^{(2)}(x) \,.
\end{equation}
For perfect symmetry (ion perfectly aligned with the mid-point in 
between the two plates), the $D$ coefficient is evaluated as 
\begin{equation}
\label{D1Z}
D(1,\calZ) =  D\left(\calZ=\frac{a}{2}\right) = \frac{12 \, \zeta(3)}{\calZ^3} =
\frac{14.4247}{\calZ^3} < 2 \times \frac{8}{a^3} \,.
\end{equation}
The contribution of the additional 
mirror charges reduces the total result 
by about 10\,\% in comparison to the 
contribution from the two closest mirror charges alone,
which would otherwise suggest a value of 
$16/a^3$ for the $D$ coefficient.

%
%
\subsection{``Sniper'' Setup Configuration}

The basic idea is obvious from Fig.~\ref{fig1}.  An ion enters the beamline,
pre-accelerated.  Energy selection (velocity selection) with a control
measurement of the ion's energy proceeds before entering the interaction region between
the two aligned, parallel plates. For an accelerating voltage of order $20\,
{\rm V}$, and single-charged helium ions (${\rm He}^+$), the de Broglie
wavelength is of the order of $5.1 \times 10^{-13}\,{\rm m}$. Thus, 
we can safely ignore diffraction effects
which could otherwise occur when entering the interaction region in between the
plates.  Interactions with the image charges on both
sides of the beam track can be ignored under perfectly aligned (``sniper'')
conditions.  The idea is that the atom loses energy on its trajectory due to
non-contact friction, with a corresponding 
measurable energy loss after its has left the interaction region.  An
aperture before the ion's energy measurement area, about a meter away, ensures
that no significant distortion of the ion's path due to interaction with the
image charges has occurred. This is necessary because the 
Coulomb interaction with the image charges does work on the projectile
ion, potentially altering its kinetic energy.

The friction force in the interaction region is given as
\begin{equation}
F_\parallel = - \eta_{(2)} \, v_x \,,
\end{equation}
where $\eta_{(2)}$ is the damping coefficient for a
configuration with two parallel plates,
which we consider according to Eq.~\eqref{matching},
within the approximation~\eqref{D1Z} for the
ion in the middle of the parallel plates.
The conductivities of metals strongly depend on the concentration 
of dopants. We use the following values, which represent
estimates of the room-temperature direct-current 
conductivity $\sigma = \sigma_T(0)$ of a number of metals which can easily 
be formed into almost perfect slabs, either as a 
polycrystalline material with a well-polished surface,
or even as single crystals (as in the case of vanadium),
\begin{subequations}
\label{metals}
\begin{align}
\sigma_{\rm Au} \approx & \; 2.30 \times 10^7 (\Omega {\rm m})^{-1} \,,
\quad
\sigma_{\rm Va} \approx 5.08 \times 10^6 (\Omega {\rm m})^{-1} \,,
\\[0.133ex]
\sigma_{\rm Ti} \approx & \; 1.27 \times 10^6 (\Omega {\rm m})^{-1} \,,
\quad
\sigma_{\rm C}  \approx 1.28 \times 10^5 (\Omega {\rm m})^{-1} \,.
\end{align}
\end{subequations}
Here, $ \sigma_{\rm C}$ refers to graphite, where we 
assume that the basal plane of the hexagonal crystal 
lattice is aligned with the surface plane of the slabs.
[The data in Eq.~\eqref{metals} has been compiled as 
the average of data given by manufacturers for 
several commercially available metals with different
polycrystalline structure and different dopants;
in a precision experiment, it would seem indicated
to measure the low-frequency (direct-current) 
limit of the conductivity $\sigma$ of a specific sample
independently.] 
For reference purposes, we here indicate
that an evaluation of $L$ according to Eqs.~\eqref{Lconduc}
and~\eqref{metals} for graphite leads to a value of
$L = 1.38 \times 10^{-16} \, ({\rm rad}/{\rm s})^{-1} $.
For helium ions at a distance $\calZ$ from the plates,
with 
\begin{equation}
m = 6.646 \times 10^{-27} \, {\rm kg} \,,
\qquad 
\calZ = 0.5 \, \mu{\rm m} \,,
\end{equation}
one has $m \, \dd v_x / \dd t = -\eta_{(2)} \, v_x$ or 
\begin{equation}
\frac{\dd v_x}{\dd t} 
= -\frac{2 \eta_{(2)}}{m} \, v_x
= -\Gamma \, v_x \,.
\end{equation}
Helium ions (He$^+$) pre-accelerated to $20 \, {\rm eV}$ energy enter
the interaction region at a speed of $3.11 \times 10^4 \; {\rm m}/{\rm s}$.
Assuming a $10 \, {\rm cm}$ long interaction region, 
the fractional loss in the flight velocity is calculated as 
\begin{equation}
r = 1 - \exp(-\Gamma \, \Delta t) \,,
\qquad
\Delta t = 3.22 \times 10^{-6} \, {\rm s} \,,
\end{equation}
where $\Delta t$ is the flight time in the interaction region.
We obtain the following fractional kinetic energy losses,
\begin{subequations}
\begin{align}
r_{\rm Au} =& \; 1.55 \times 10^{-7} \,,
\quad
r_{\rm Va} = 7.03 \times 10^{-7} \,,
\\[0.133ex]
r_{\rm Ti} =& \; 2.82 \times 10^{-6} \,,
\quad
r_{\rm C} = 2.80 \times 10^{-5} \,.
\end{align}
\end{subequations}
The corresponding relative decrease in the kinetic energy 
(proportional to the square of the velocity)
is twice as large as these values.
The ``sniper'' aspect of the configuration sketched
in Fig.~\ref{fig1} comes into play when we restrict,
geometrically, the acceptance region for 
the ions leaving the interaction region.
Let us assume that roughly $D = 1\,{\rm m}$ further down 
the beam line, we  restrict the 
available angular region for the arriving ions to 
a circular aperture of diameter (see also Fig.~\ref{fig1})
\begin{equation}
a_s = 100 \, \calZ = 50 \, \mu{\rm m} \,.
\end{equation}
For non-perfect alignment, the image charge interaction
may exert a force in the interaction region,
in the positive or negative $z$ direction, 
without altering the $x$ component $v_x$.
The calculation of the interaction energy
with the mirror charge(s) has been discussed in Sec.~\ref{mirror}.
However, in order to obtain a figure-of-merit for 
our proposed experimental setup, it is not sufficient
to consider the geometric requirement of
passage through the geometric aperture indicated in Fig.~\ref{fig1};
it corresponds to a restriction of the 
modulus of the velocity change according to 
\begin{equation}
v_x \, \sqrt{ 1 + \left( \frac{a_s}{D} \right)^2 } - v_x 
\approx \frac12 \, \left( \frac{a_s}{D} \right)^2 \, v_x = \xi \, v_x \,,
\end{equation}
where we redefined $\xi$ as compared to Eq.~\eqref{defxi}.
The figure-of-merit of the measurement is obtained
as the ratio of the relative change in the velocity 
due to non-contact friction, divided by the geometrically 
restricted ``uncertainty'' in the velocity measurement,
\begin{equation}
f = \frac{r}{\xi} = 
\frac{2 \, (1 - \exp(-\Gamma \, \Delta t))}{(a_s/D)^2} \,.
\end{equation}
We obtain
\begin{subequations}
\begin{align}
f_{\rm Au} =& \; 1.24 \times 10^2\,,
\quad
f_{\rm Va} = 5.62 \times 10^2 \,,
\\[0.133ex]
f_{\rm Ti} =& \; 2.25 \times 10^3 \,,
\quad
f_{\rm C} = 2.23 \times 10^4 \,.
\end{align}
\end{subequations}
These results show that a measurement should be feasible.
We also have done an evaluation based on the extensive
reference volume~\cite{Pa1985}, analyzing available 
data for the dielectric function of $\alpha$-quartz,
along the ordinary (o) and extraordinary (e) axis.
We obtain the following values for the $L$ coefficient,
\begin{subequations}
\begin{align}
L_o =& \; 1.40 \times 10^{-17} \, ({\rm rad}/{\rm s})^{-1} \,, 
\\[0.133ex]
L_e =& \; 2.14 \times 10^{-17} \, ({\rm rad}/{\rm s})^{-1} \,,
\end{align}
\end{subequations}
which translates into the following loss coefficients and 
figures-of-merit,
\begin{subequations}
\begin{align}
r_o &=  2.82 \times 10^{-6} \,,
\qquad &
r_e & =  4.30 \times 10^{-6} \,,
\\[0.133ex]
f_o & = 9.04 \times 10^3 \,,
\qquad &
f_e & = 1.37 \times 10^4 \,.
\end{align}
\end{subequations}
A measurement using single-crystalline quartz 
surfaces also seems possible.
The ``sniper'' configuration is ``auto-correcting''
in the sense that the ions can only 
pass through the detector aperture under well-aligned conditions.

%
%
\section{Conclusions}
\label{sec5}

We investigate the energy loss of an ion in the vicinity of two
conducting surfaces, in a configuration where the electrostatic interaction
with the two image charges compensate each other.  Based on a rederivation of
the frictional force, we relate the general expression for the friction force
to the functional form of the permittivity at low frequencies, which we
parameterize in terms of a sum of generalized Drude and Lorentz profiles.  The
definition of a conductor implies the existence of a zero-resonance-frequency
term in the permittivity and we show that it is this term which dominates in
the evaluation of the frictional force.

Specifically, the result for the non-contact friction coefficient, given in
Eq.~\eqref{ion_general} for a single wall, is generalized to the interaction
with two parallel walls in Eq.~\eqref{matching}.  The $L$ coefficient can be
written in terms of a parameterization of the dielectric response functions of
the material, according to Eq.~\eqref{epsmod}.  For a conductor, the limit $L$
is exclusively determined by the conduction band [see Eq.~\eqref{Lconduc}].

We identify the main problem in a conceivable experiment as the isolation of
the frictional force from the strong electrostatic interaction with the image
charge. In general, the functional form of the interaction potentials of ions
with other electromagnetically interacting objects is different from the
corresponding expressions for atoms.  In Sec.~\ref{sec1}, we 
recall that the ion-atom interaction ($1/R^4$) is stronger than the
atom-atom interaction ($1/R^6$), and that the friction force in the vicinity of
a surface ($1/\calZ^3$) is larger than that for an atom ($1/\calZ^5$).  In a
typical case, there is a difference in the power law involving two more inverse
powers of the distance for ``atom-something'' interactions as compared to the
corresponding ``ion-something'' interaction.

In principle, the enhanced functional form of the non-contact friction coefficient
for ions as compared to atoms would recommend a measurement of the quantum
friction effect using ions. However, the strong interaction with the image
charge implies that a careful compensation of the interactions with the images
becomes necessary in the vicinity of a conducting surface, in order to
minimize a deflection of the ion's trajectory.  We aim to identify a
collection of suitable parameters, in terms of a pre-accelerating voltage, to
find a compromise between a slow flight velocity (which enhances the ``risk of
deflection'') and a too short interaction time (which would make the quantum
friction effect undetectable).  A proposed set-up with certain auto-compensating
features, to isolate the friction effect, is described in Sec.~\ref{sec4}.  The
most suitable materials for the experiment are be those which are readily
available as large-scale single crystals, and constitute conductors, but with a
low value of the conductivity, in order to increase the Ohmic heating due to
the flow of the image charge.  We find that
the best figure-of-merit is obtained for graphite, while a readily available
metal like vanadium, which has excellent surface properties, also would be
available for a precision experiment. 

%
%
\section*{Acknowledgments}

The authors acknowledge helpful conversations with Professor K.~Pachucki.  This
research has been supported by the National Science Foundation 
(Grants PHY--1068547 and PHY--1403973)
and by the Polish Ministry of Science (MNiSW,
Grant No. 0307/IP3/2011/71).
Early stages of this research have also been supported by the 
Deutsche Forschungsmeinschaft (DFG, contract Je285/5--1).

\appendix

%
%
\section{Fourier Transform in Signal Theory}
\label{appa}

In signal theory, different conventions are used for the Fourier transform.
Normally, in physics, one integrates the Fourier frequency transform with the
``integration measure'' $\dd \omega/(2 \pi)$.  However, according to the
Eq.~(3.118) of Ref.~\cite{RyKrTa1989vol2}, and Eqs.~(1.90) and (1.91) of
Ref.~\cite{RyKrTa1989vol3}, we have
\begin{subequations}
\begin{align}
f(t) =& \; \int \dd \omega \, \ee^{-\ii \, \omega \, t} \, 
{\widetilde f}(\omega) \,,
\\[0.133ex]
{\widetilde f}(\omega) 
=& \; \int \frac{\dd t}{2 \pi} \, \ee^{\ii \, \omega \, t} \, f(t) \,,
\end{align}
\end{subequations}
and the same conventions are employed in signal theory 
for the corresponding transformations from 
coordinate space to wave number space,
namely, one integrates with the integration measures
$\dd^3 r/(2\pi)^3$ and $\dd^3 k$ instead of
$\dd^3 r$ and $\dd^3 k/(2\pi)^3$.
The difference in the integration measures
also explains the occurrence of multiplicative 
factors of $2 \pi$ in other formulations of the 
fluctuation-dissipation theorem,
such as Eq.~(3.118) of Ref.~\cite{RyKrTa1989vol2}.

\end{document}